\documentclass[aps,prc,twocolumn,nofootinbib,showpacs,floatfix]{revtex4}
\usepackage{epsfig}
\usepackage{bm}
\usepackage{dcolumn}

\usepackage[normalem]{ulem}  
\usepackage[dvips]{color} 

\def\bea {\begin{eqnarray}}
\def\eea {\end{eqnarray}}
\def\be {\begin{equation}}
\def\ee {\end{equation}}
\def\ben{\begin{enumerate}}
\def\een{\end{enumerate}}
\def\bi{\begin{itemize}}
\def\ei{\end{itemize}}

\def\O{{\cal O}}
\def\F{{\cal F}}

\def\prl {Phys. Rev. Lett.\ }
\def\pl {Phys. Lett.\ }
\def\pr {Phys. Rev.\ }
\def\np {Nucl. Phys.\ }

\def\GV{G_{\mbox{\tiny V}}}
\def\GF{G_{\mbox{\tiny F}}}
\def\DRV{\Delta_{\mbox{\tiny R}}^{\mbox{\tiny V}}}
\def\hyphen{{\mbox{-}}}

\newcommand{\sfrac}[2]{\mbox{\small{$\frac{#1}{#2}$}}}

\def\2p{|2p\rangle }
\def\4p2h{|4p\hyphen 2h\rangle }
\def\6p4h{|6p\hyphen 4h\rangle }
\def\2h{|2h\rangle }
\def\4h2p{|4h\hyphen 2p\rangle }
\def\6h4p{|6h\hyphen 4p\rangle }

\begin{document} 
\preprint{ }
 
\title{Comparative tests of isospin-symmetry-breaking corrections
to superallowed $0^{+}$$\rightarrow$$0^{+}$ nuclear $\beta$ decay}

\author{I.S. Towner}
\email[]{towner@comp.tamu.edu}
\author{J.C. Hardy}
\email[]{hardy@comp.tamu.edu}
\affiliation{Cyclotron Institute, Texas A\&M University,                    
College Station, Texas  77843}
\date{\today} 
\begin{abstract} 

We present a test with which to evaluate the calculated isospin-symmetry-breaking corrections
to superallowed $0^{+} \rightarrow 0^{+}$ nuclear $\beta$ decay.  The test is based on the
corrected experimental $\F t$ values being required to satisfy conservation of the vector current
(CVC).  When applied to six sets of published calculations, the test demonstrates quantitatively
that only one set -- the one based on the shell model with Saxon-Woods radial wave functions --
provides satisfactory agreement with CVC.  This test can easily be applied to any sets of calculated
correction terms that are produced in future.      

\end{abstract} 


\maketitle

\section{Introduction}
\label{s:intro}

Superallowed $0^{+} \rightarrow 0^{+}$ $\beta$ decay between $T=1$ nuclear analog
states has been a subject of continuous and often intense study for six decades.  The
$ft$ values for such transitions are nearly independent of nuclear-structure ambiguities
and depend uniquely on the vector part of the weak interaction.  Their measurement gives
us access to clean tests of some of the fundamental precepts of weak-interaction theory,
and, over the years, this strong motivation has led to very high precision being achieved
both in the experiments and in the theory used to interpret them.

The most recent survey of world data \cite{HT09} finds ten of these superallowed transitions
with measured $ft$ values known to 0.1\% precision or better, and three more that have a precision
of between 0.1 and 0.3\%.  An analysis of the $ft$ values \cite{HT09} demonstrated that
the vector coupling constant, $\GV$, has the same value for all thirteen transitions to within
$\pm$0.013\%, thus confirming a key part of the Conserved Vector Current (CVC) hypothesis; and it
set an upper limit on a possible scalar current at 0.2\% of the vector current.  With both these
outcomes established, the results could then be used to extract a value for $V_{ud}$, the up-down
element of the Cabibbo-Kobayashi-Maskawa (CKM) matrix, with which the top-row unitarity test of
the CKM matrix yielded the result \cite{TH10} 0.9999(6).  This is in remarkable agreement with the
Standard Model, and the tight uncertainty significantly limits the scope for any new physics beyond
the model.  Further tightening of the uncertainty would increase the impact of this result even more.

Although the role played by nuclear structure is relatively small, the precision currently reached
by experiment is such that the theoretical uncertainties introduced by correction terms required in
the analysis of the $ft$-value data now predominate over the experimental uncertainties.  Two of
these correction terms depend on nuclear structure and together they are the second largest contributor
to the overall uncertainty in $V_{ud}$.  The largest contributor is the nucleus-independent component
of the radiative correction but at present there seems little opportunity for further improvement there.

Thus it is the nuclear-structure dependent terms that have attracted the greatest attention, particularly
recently.  The most widely used of these latter correction terms are those calculated by the present
authors, which have been tabulated for all the superallowed transitions of interest in Ref.~\cite{TH08}.  
However, there are a growing number of alternative choices \cite{Da69,OB85,OB95,SGS96,LGM09,Au09} available
for one of the two correction terms -- the one that accounts for isospin symmetry breaking -- including
a set we offer ourselves \cite{HT09}.  There has also been a claim, albeit unsupported by any
detailed computations, that our calculations neglect a radial excitation term, which is purported to
be important \cite{MS09}.  To counterbalance that, however, there are two recent papers that confirm
our result: one \cite{Gr10} does so based on a semi-empirical analysis of the data, while the other
\cite{Sa10} quotes the average results from a Skyrme-density-functional-theory calculation in which
simultaneous isospin and angular-momentum projection has been incorporated. 

Clearly it would be valuable if the various sets of calculated isospin-symmetry-breaking correction terms
could be tested against the data, and their relative merits quantitatively evaluated, since this must
surely be a first step in any attempt to reduce the uncertainty attributed to these corrections.  In
this paper, we address ourselves to devising and then applying such a test.

We begin by describing how information on the fundamental weak-interaction parameters is extracted
from the experimental $ft$-value data.  We will overview the role played by all the theoretical
corrections but will focus, in particular, on the isospin-symmetry-breaking term.  This will lead
naturally to the test we propose as a means of evaluating the efficacy of any calculated set of these
terms available now or in the future.  We will then outline the methods currently used to calculate
the isospin-symmetry-breaking term, and proceed to apply our test to each of them.  Finally we will
evaluate the results of the test and present our conclusions.

\section{The analysis of superallowed beta transitions}
\label{s:sbd}

Superallowed Fermi beta decay between $0^+$ states depends uniquely on the vector part of the hadronic
weak interaction.  According to CVC, when the decay occurs between isospin $T=1$ analog states the
measured $ft$ values should be the same irrespective of the nucleus, {\it viz.}
\be
ft = \frac{K}{\GV^2 | M_F |^2} = {\rm ~const},
\label{ftconst}
\ee
where $K/(\hbar c )^6 = 2 \pi^3 \hbar \ln 2 / (m_e c^2)^5 = ( 8120.2787 \pm 0.0011 ) \times 10^{-10}$
GeV$^{-4}$s; $\GV $ is the vector coupling constant for semi-leptonic weak interactions; and $M_F$ is
the Fermi matrix element.  The CVC hypothesis asserts that the vector coupling constant, $\GV$, is a
true constant and not renormalised to another value in the nuclear medium.

In practice, Eq.~(\ref{ftconst}) has to be amended slightly.  Firstly, there are radiative corrections
because, for example, the emitted electron may emit a bremsstrahlung photon that goes undetected in the
experiment.  Secondly, isospin is not an exact symmetry in nuclei so the nuclear matrix element, $M_F$, 
is not the same for all superallowed transitions but is slightly reduced from its ideal value by a
different amount in each case.  This leads us to write:
\be
|M_F|^2 = |M_0|^2 ( 1 - \delta_C ) ,
\label{MF2}
\ee
where $M_0$ is the exact-symmetry value, which for $T = 1$ states is $M_0 = \sqrt{2}$; and $\delta_C$ is
the isospin-symmetry-breaking correction, which takes on a different (small) value for each transition.  
Thus, we define a ``corrected" $\F t$ value as
\be
\F t \equiv ft (1 + \delta_R^{\prime}) (1 + \delta_{NS} - \delta_C ) =
\frac{K}{2 \GV^2(1 + \DRV )},
\label{Ftfactor}
\ee
where, in addition to the terms already defined, $\DRV$ is the transition-independent part of the
radiative correction; while the terms $\delta_R^{\prime}$ and $\delta_{NS}$ comprise the
transition-dependent part of the radiative correction, the former being a function only of the electron's
energy and the $Z$ of the daughter nucleus, while the latter, like $\delta_C$, depends in its evaluation
on the details of nuclear structure.

From this equation, it can be seen that a single measured transition establishes a value for $\F t$, and
hence $\GV$.  This result could, in principle, then be used to determine $V_{ud}$ via the relationship
$V_{ud}$ = $\GV/\GF$, where $\GF$ is the well known weak-interaction constant for muon decay \cite{PDG08}.  
However, a value for $V_{ud}$ derived from a single superallowed transition would be reliant upon a single
pair of structure-dependent correction terms, $\delta_{NS}$ and $\delta_C$, without there being any independent
verification of those terms' validity; so, in practice, as many transitions as possible are measured and
their resultant $\F t$ values compared.  If they satisfy CVC by being statistically consistent with each
another, then one is justified in taking an average value of $\F t$, from which $\GV$ and $V_{ud}$ can then
be derived.

If they are not consistent with each other, then one can proceed no further since inconsistency must
signal a failure either of the calculated structure-dependent corrections or else of the CVC hypothesis
itself.  In either case, an average value of $\F t$ has no defined significance and certainly cannot be
used to obtain a value for $V_{ud}$.

Here we find the basis for a test of the calculated structure-dependent correction terms: How well do
they do in producing a consistent set of $\F t$ values from the experimental $ft$ values?  The latter show
very pronounced differences from one transition to another, and the extent to which those differences are
successfully removed by a given set of calculated correction terms would be a sensitive measure of the
efficacy of the calculations involved.  Naturally, such a test is only as good as the CVC hypothesis.  
However, we believe that most would agree that a persistent scatter in the derived $\F t$ values is more
likely to be due to a deficiency in the calculated corrections rather than to a failure of CVC.

\section{The test}
\label{s:test}

Our test is based upon the premise that CVC is valid at least to $\pm$0.03\%, which is the level of
precision currently attained by the best $ft$-value measurements.  Under that condition, a valid set of
structure-dependent correction terms should produce a statistically consistent set of $\F t$ values, the
average of which we can write as $\overline{\F t}$.  It then follows from Eq.~(\ref{Ftfactor}) that, for
each individual transition in the set, we can write
\be
\delta_C - \delta_{NS} = 1 - \frac{\overline{\F t}}{ft (1 + \delta_R^{\prime})}.
\label{test}
\ee
For any set of corrections to be acceptable, the calculated value of $\delta_C - \delta_{NS}$ for each
superallowed transition must satisfy this equation, where $ft$ is the measured result for that transition and
$\overline{\F t}$ has the same value for all of them.  Thus, to test a set of correction terms for $n$
superallowed transitions, one can treat $\overline{\F t}$ as a single adjustable parameter and use it
to bring the $n$ results from the right side of Eq.~(\ref{test}), which are based predominantly on experiment, 
into the best possible agreement with the corresponding $n$ calculated values for $\delta_C - \delta_{NS}$.
The normalized $\chi^2$, minimized by this process, then provides a figure of merit for that set of
calculations. 

As it happens, there is only one set of calculations available for $\delta_{NS}$ \cite{To94,TH08} but
many for the isospin-symmetry-breaking term $\delta_C$.  It therefore becomes more useful to rearrange
Eq.~(\ref{test}) to read:
\be
\delta_C = 1 + \delta_{NS} - \frac{\overline{\F t}}{ft (1 + \delta_R^{\prime})}.
\label{finaltest}
\ee 
The same least-squares minimization process can of course be used in the application of this equation.

\section{Available calculations for $\delta_C$}
\label{s:calc}

There have been a number of methods used over the years to calculate the isospin-symmetry-breaking
correction to superallowed $\beta$ decay.  We describe some of them here, in chronological order.

\subsection{Damgaard model}
\label{ss:Dam}

The first model was proposed in 1969 by Damgaard \cite{Da69} and was improved 8 years later by
Towner, Hardy and Harvey \cite{THH77}.  The idea is that the proton involved in beta decay has a
different radial wave function than the neutron into which it transforms because it is influenced
by the Coulomb interaction with all the other protons in the nucleus.  If the other protons present
a uniform charge distribution of radius, $R$, then the Coulomb interaction for a proton at $r<R$ is
\be
V_c(r) = - \frac{Z e^2}{R^3} \sum_{i = 1}^A \left ( \sfrac{1}{2} r_i^2 -
\sfrac{3}{2} R^2 \right ) \left ( \sfrac{1}{2} - t_z(i) \right )
\delta ( r - r_i),
\label{Vc1}
\ee
where $t_z(i) = - \sfrac{1}{2}$ if nucleon $i$ is a proton, and $ = + \sfrac{1}{2}$ if it is a neutron. 

Using an oscillator model as a basis, Damgaard expanded the proton radial function in terms of a complete
set of neutron oscillator functions.  The set comprised states of the same orbital angular momentum,
$\ell$, but differing numbers of radial nodes, $n$.  Most of the mixing turned out to be with the state
with one more radial node, so
\be
u_{\ell}^{proton}(r) \approx (1-\alpha^2)^{1/2} u_{n,\ell}(r) + \alpha u_{n+1,\ell}(r).
\label{oscexp}
\ee
The mixing amplitude comes from first-order perturbation theory,
\bea
\alpha & = & \langle u_{n+1,\ell}|V_c|u_{n,\ell}\rangle (\Delta E)^{-1}
\nonumber \\
\Delta E & = & E_{n+1,\ell} - E_{n,\ell} = 2 \hbar \omega,
\label{Damgalpha}
\eea
and the Fermi matrix element between $T=1$ states is given by
\be
|M_F|^2 = 2 (1 - \alpha^2).
\label{MFDamg}
\ee

Upon evaluating the Coulomb matrix element using oscillator functions with $V_c$ taken from
Eq.~(\ref{Vc1}), Damgaard obtained
\be
\delta_C = \alpha^2 = \frac{Z^2}{(\hbar \omega)^4 R^6} \frac{e^4 \hbar^4}{
16 m^2} (n+1)(n+\ell + \sfrac{3}{2}).
\label{dcDamg1}
\ee
If we adopt the relationships $\hbar \omega = 41 A^{-1/3}$ MeV and $R = 1.2 A^{1/3}$ fm, this
expression becomes
\be
\delta_C = 0.2645 \times Z^2 A^{-2/3} (n+1) (n+\ell + \sfrac{3}{2}),
\label{dcDamg2}
\ee
which, for the light nuclei we are interested in, exhibits the general behaviour $\delta_C \propto A^{4/3}$
with some shell structure superimposed through the choice of oscillator quantum numbers $n$ and $\ell$.  In
particular, a proton radial function with one radial node gets a factor of two enhancement in its $\delta_C$
value over one that has no radial nodes simply from the factor $(n+1)$ in Eq.~(\ref{dcDamg2}).

We have used Eq.~(\ref{dcDamg2}) to derive $\delta_C$ values for the thirteen best known superallowed
transitions.  These transitions are listed by parent nucleus in the first column of Table~\ref{t:dc}, and the
$\delta_C$ results for this model appear in the fifth column of the same table.  We will use these results in
our comparative tests of all models.

\subsection{Shell model with Saxon-Woods radial wave functions (SM-SW)}
\label{ss:SMSW}

This model was introduced by Towner, Hardy and Harvey in 1977 \cite{THH77} and improved upon several times
since then \cite{TH02,TH08}.  In their approach, the Fermi matrix element is defined by
\be
M_F = \sum_{\alpha} \langle f|a_{\alpha}^{\dag} b_{\alpha} | i \rangle = 
\sum_{\alpha ,\pi} \langle f|a_{\alpha}^{\dag}|\pi \rangle
\langle \pi|b_{\alpha}|i\rangle ,
\label{MF}
\ee
where $a_{\alpha}^{\dag}$ creates a neutron and $b_{\alpha}$ annhilates a proton in state $\alpha$.  Here
$|i \rangle$ and $|f \rangle$ are the {\it exact} $A$-body state vectors for the full Hamiltonian, and
$|\pi \rangle$ represents a complete set of ($A-1$)-body parent states.  If this Hamiltonian commutes
with the isospin operators, then $|i \rangle$ and $|f \rangle$ are exact isospin analogues of each other, 
and the symmetry-limit matrix element is 
\be 
M_0 = \sum_{\alpha , \pi} | \langle f|a_{\alpha}^{\dag} | \pi \rangle |^2,
\label{M0}
\ee
which for $T = 1$ states corresponds to $M_0 = \sqrt{2}$.  However, with isospin not being an exact
symmetry, $|i\rangle$ and $|f\rangle$ are not exact isospin analogues; nevertheless the resulting
matrix element $M_F$ is not very different from $M_0$, the relationship between them being given by
Eq.~(\ref{MF2}): {\it viz.} $M_F^2 = M_0^2 (1 - \delta_C)$, where $\delta_C$ is small.

Ideally, to obtain
$\delta_C$ one would compute Eq.~(\ref{MF}) using the shell model, and introduce Coulomb
and other charge-dependent terms into the shell-model Hamiltonian.  However, the shell-model space
would have to be huge to include all the potential states with which the Coulomb interaction
might potentially connect.  Since this is not a practical proposition, a model approach was
developed in which $\delta_C$ is divided into two parts:
\be
\delta_C = \delta_{C1} + \delta_{C2}.
\label{dc1dc2}
\ee

For $\delta_{C1}$, one computes
\be
\sum_{\alpha ,\pi} \langle \overline{f}|a_{\alpha}^{\dag}|\pi \rangle
\langle \pi |b_{\alpha}| \overline{\imath}\rangle =
M_0 (1 - \delta_{C1})^{1/2},
\label{dc1}
\ee
where $|\overline{\imath}\rangle$ and $|\overline{f}\rangle$ are not the exact eigenstates that appear in
Eq.~(\ref{MF}) but are the shell-model eigenstates of an effective Hamiltonian (including charge-dependent
terms) evaluated in a tractable shell-model space.  However, this space is not large enough to allow for mixing with
functions having a different number of radial nodes, so the term $\delta_{C2}$ is introduced to compensate
for that limitation.  This second term is derived from
\be
\sum_{\alpha ,\pi} |\langle \overline{f}|a_{\alpha}^{\dag}|\pi \rangle |^2
r_{\alpha}^{\pi} = M_0 (1 - \delta_{C2})^{1/2},
\label{dc2}
\ee
where $r_{\alpha}^{\pi}$ is a radial overlap integral of proton and neutron radial functions.  If the proton
and neutron radial functions were identical, then it would follow that $r_{\alpha}^{\pi} = 1$, and $\delta_{C2}
=0$.  But, since they are not identical, a finite correction $\delta_{C2}$ is obtained.  The idea is that nodal
mixing mainly impacts on the radial functions -- as demonstrated by the Damgaard model -- and so its impact is
best modelled by Eq.~(\ref{dc2}).

The wave function for the decaying $A$-body state, $|\overline{\imath}\rangle$, is expanded in a set of parent
states of $(A-1)$-nucleons, $|\pi \rangle$, plus a proton; while that of the daughter $A$-body state,
$|\overline{f}\rangle$, is expanded in terms of the same set of parent states plus a neutron.  The expansion
coefficients are obtained from a shell-model calculation.  Isospin-symmetry breaking is introduced by allowing
the radial function for the proton in these expansions to differ from that of the neutron.  In this model,
these radial functions are taken to be eigenfunctions of a Saxon-Woods potential.  The well-depths of the
proton and neutron potentials are adjusted so that the asymptotic forms of the radial function go as
$e^{- \alpha r}$, where $\alpha^2 = 2 m S / \hbar^2$, with $m$ being the nucleon mass, and $S$ the experimental
separation energy for the proton (or neutron) in the $A$-body state.  Further details can be found in \cite{TH08}.

It is important to realize that this model is really semi-phenomenological in its application.  In addition to
the match with experimental separation energies in the calculation of $\delta_{C2}$, the radius of the Saxon-Woods
potential in each case was set to the value determined experimentally for the charge radius by electron
scattering \cite{TH02}, and the shell-model parentage was linked to measured single-nucleon transfer reactions
\cite{TH08}.  The value of $\delta_{C1}$ was also constrained by comparison with experiment.  First, for each
superallowed transition the single-particle energies of the proton orbits were shifted relative to the neutrons, 
the exact amount being determined from the spectrum of single-particle states in the closed-shell-plus-proton
versus the closed-shell-plus-neutron nucleus.  Second, the two-body Coulomb interaction among the valence protons
was adjusted in strength for each decay so that the measured $b$ coefficient in the isobaric multiplet mass
equation (IMME) was exactly reproduced for the multiplet involved in that decay.  Third, the charge-dependent
nuclear interaction, which had been incorporated by a $\sim$2\% increase in all the $T = 1$ proton-neutron matrix
elements relative to the neutron-neutron ones, was tuned to give agreement with the measured $c$ coefficient of
the IMME.

The current best values for $\delta_C$ as calculated with this model are listed in the fifth column of Table VI
in Ref.~\cite{TH08}, and are reproduced here in the sixth column of Table~\ref{t:dc}.

\subsection{Shell model with Hartree-Fock radial wave functions (SM-HF)}
\label{ss:SMHF}
 
Beginning in 1985, Ormand and Brown \cite{OB85,OB95} adopted the same general procedure as the one just described, 
splitting $\delta_C$ into two components, the first of which, $\delta_{C1}$, incorporated configuration mixing
within a restricted shell-model space, and the second, $\delta_{C2}$, accounted for mixing with
all other states by evaluating the mismatch in the parent and daughter radial wave functions.  The shell-model
aspects of their model were the same as the SM-SW model, but their radial functions were taken to be
eigenfunctions of a mean-field Hartree-Fock potential rather than of a Saxon-Woods potential.  As in the SM-SW
model, the strength of this mean field was readjusted so that the asymptotic forms of the proton and neutron
radial functions were matched to their respective separation energies.

Ormand and Brown's \cite{OB85,OB95} protocol was to perform two Hartree-Fock calculations with a Skyrme interaction: 
one for the decaying $A$-body state, whose mean field provided the proton function, and the other for the
daughter $A$-body state, whose mean field provided the neutron function.  However, it was noted more recently
by Hardy and Towner \cite{HT09} that there is a problem with this protocol: the Coulomb part of the proton mean
field has asymptotically the wrong form, falling off as $(Z+1)e^2/r$ rather than $Z e^2/r$.  They therefore
modified the protocol to just a single Skyrme-Hartree-Fock calculation performed in the $(A-1)$-body state, whose mean
field provided for both the proton and neutron radial functions.  In this procedure, the Coulomb interaction
automatically has the right asymptotic form.  Further details can be found in \cite{HT09}.

We have obtained the $\delta_C$ values for this model by adding the ``adopted" $\delta_{C1}$ numbers from Table
III of Ref.~\cite{TH08} (the same as we used for the SM-SW model) and the ``HF" $\delta_{C2}$
numbers from Table XI of Ref.~\cite{HT09}.  The results appear in column seven of our Table~\ref{t:dc}.

\subsection{Hartree-Fock with Random Phase Approximation (RHF-RPA and RH-RPA)}
\label{ss:HFRPA}

In 1996, Sagawa, Van Giai and Suzuki \cite{SGS96} introduced a new model, in which a Skyrme-Hartree-Fock calculation
was performed for each even-even $A$-body system: the parent for the cases in which the superallowed decay proceeds
from a $T_z = -1$ parent nucleus, and the daughter for cases of decay from a $T_z = 0$ parent nucleus.  The odd-odd
nucleus was then treated as a particle-hole excitation built on the even-even Hartree-Fock state.  The particle-hole
calculation was carried out in the charge-exchange random-phase approximation (RPA) in a model space extending up to
$10 \hbar \omega$ excitation, with radial functions up to five nodes.  The lowest state in the RPA spectrum was
identified as the isobaric analogue state -- the state actually involved in the superallowed Fermi beta decay.  
Unlike the previous two methods, there was no adjustment to reproduce exactly the energy of the analog state, but
the authors did check that their results were typically within 500 keV of the experimental value.  Isospin-symmetry
breaking was introduced by the presence of a Coulomb interaction, augmented by explicit charge-symmetry-breaking
and charge-independence-breaking interactions included in the two-body force used in the Hartree-Fock calculation.

\begin{table*}[t]
\begin{center}
\caption{Input data for the tests of the isospin-symmetry-breaking corrections, $\delta_{C}$, obtained from the
various models described in Sect.~\ref{s:calc}.  The experimental $ft$ values come from Table IX in the most recent
survey of world data \cite{HT09}; however, in order to ensure undiluted normal statistics, we have set all the ``scale
factors" used in that reference equal to 1, with the consequence that the uncertainties quoted for most cases are
smaller than those listed in Ref.~\cite{HT09}.  The calculated values of $\delta^{\prime}_R$ and $\delta_{NS}$ come from
Table VII of Ref.~\cite{TH08}.  The $\delta_C$ values tabulated in the last six columns were obtained as follows:
the ``Damgaard" values were derived from our Eq.~(\ref{dcDamg2}); those labeled SM-SW came from Table VII of
Ref.~\cite{TH08}; the SM-HF values were obtained by adding the ``adopted" $\delta_{C1}$ numbers from Table III of
Ref.~\cite{TH08} and the ``HF" $\delta_{C2}$ numbers from Table XI of Ref.~\cite{HT09}; the RHF-RPA values were
taken for the PKO1 effective interaction given in Table I of Ref.~\cite{LGM09}; the RH-RPA numbers, which
corresponded to the density-dependent DD-ME2 effective interaction, were taken from the same table and reference;
the IVMR values were calculated from our Eq.~(\ref{IVMdc2}), which is the same as Eq.~(32) in Ref.~\cite{Au09}.
\label{t:dc}}
\vskip 1mm
\begin{ruledtabular}
\begin{tabular}{llllcccccc}
& & & & & & & \\[-3mm]
\multicolumn{1}{l}{Parent} &
\multicolumn{1}{l}{Experimental} & & &
\multicolumn{6}{c}{$\delta_C$ (\%)} \\
\cline{5-10} \\[-3mm]
\multicolumn{1}{l}{nucleus} &
\multicolumn{1}{l}{$ft$ value (s)} &
\multicolumn{1}{l}{$\delta^{\prime}_R$ (\%)} &
\multicolumn{1}{l}{$\delta_{NS}$ (\%)} &
\multicolumn{1}{c}{Damgaard} &
\multicolumn{1}{c}{SM-SW} &
\multicolumn{1}{c}{SM-HF} &
\multicolumn{1}{c}{RHF-RPA} &
\multicolumn{1}{c}{RH-RPA} &
\multicolumn{1}{c}{IVMR} \\[1mm]
\hline
& & & & & & & & & \\[-3mm]
\mbox{\boldmath $T_z = -1:$} & &  && & & & & \\[1mm]
$^{10}$C  & 3041.7(43) & 1.679(4)  & -0.345(35) & 0.046 & 0.175 & 0.225 & 0.082 & 0.150 & 0.008  \\
$^{14}$O  & 3042.3(11) & 1.543(8)  & -0.245(50) & 0.111 & 0.330 & 0.310 & 0.114 & 0.197 & 0.015  \\
$^{22}$Mg & 3052.0(70) & 1.466(17) & -0.225(20) & 0.153 & 0.380 & 0.260 &       &       & 0.031  \\
$^{34}$Ar & 3052.7(82) & 1.412(35) & -0.180(15) & 0.285 & 0.665 & 0.540 & 0.268 & 0.376 & 0.064  \\[2mm]
\mbox{\boldmath $T_z = 0:$} & & & & & & & & \\[1mm]
$^{26}$Al & 3036.9(9)  & 1.478(20) & ~0.005(20) & 0.182 & 0.310 & 0.440 & 0.139 & 0.198 & 0.041  \\
$^{34}$Cl & 3049.4(11) & 1.443(32) & -0.085(15) & 0.326 & 0.650 & 0.695 & 0.234 & 0.307 & 0.064  \\
$^{38}$K  & 3051.9(5) & 1.440(39) & -0.100(15) & 0.370 & 0.655 & 0.745 & 0.278 & 0.371 & 0.077  \\
$^{42}$Sc & 3047.6(12) & 1.453(47) & ~0.035(20) & 0.414 & 0.665 & 0.640 & 0.333 & 0.448 & 0.091  \\
$^{46}$V  & 3049.5(8)  & 1.445(54) & -0.035(10) & 0.524 & 0.620 & 0.600 &       &       & 0.106  \\
$^{50}$Mn & 3048.4(7) & 1.444(62) & -0.040(10) & 0.550 & 0.655 & 0.620 &       &       & 0.122  \\
$^{54}$Co & 3050.8(10) & 1.443(71) & -0.035(10) & 0.613 & 0.770 & 0.685 & 0.319 & 0.393 & 0.139  \\
$^{62}$Ga & 3074.1(11) & 1.459(87) & -0.045(20) & 1.339 & 1.48  & 1.21  &       &       & 0.175  \\
$^{74}$Rb & 3084.9(77) & 1.50(12)  & -0.075(30) & 1.422 & 1.63  & 1.42  & 1.088 & 1.258 & 0.235  \\[1mm]
\hline \\[-2mm]
\multicolumn{4}{l}{$\chi^2/n_d$ (statistical experimental uncertainties only)} & 8.3 & 1.2 & 8.3 & 7.2 & 6.0 & 48  \\[1mm]
\multicolumn{3}{l}{Confidence level (\%)} & & 0 & 26 & 0 & 0 & 0 & 0  \\[1mm]
\hline \\[-2mm]
\multicolumn{4}{l}{$\chi^2/n_d$ (uncertainties on experiment, $\delta^{\prime}_R$ and $\delta_{NS}$)} & 1.7 & 0.4
& 2.2 & 2.7 & 2.1 & 11  \\[1mm]
\multicolumn{4}{l}{$\chi^2/n_d$ (uncertainties on experiment, $\delta^{\prime}_R$, $\delta_{NS}$ and $\delta_C$)}
& 0.9 & 0.3 & 1.1 & 1.6 & 1.3 & 4.5  \\[1mm]
\end{tabular}
\end{ruledtabular}
\end{center}
\end{table*}

Since first results from this model appeared \cite{SGS96} significant progress has been made in self-consistent RPA calculations
in charge-exchange channels.  Skyrme zero-range interactions have been replaced by finite-range meson-exchange potentials
involving $\sigma$, $\omega$, $\rho$, and $\pi$ mesons, and a relativistic rather than a non-relativistic treatment
can be used.  In 2009 Liang, van Giai and Meng \cite{LGM09} published improved results from relativistic Hartree-Fock,
RHF-RPA, calculations with three different effective interactions, as well as from relativistic Hartree (only), RH-RPA,
calculations with density-dependent meson-nucleon couplings and non-local interactions.  The results were not
particularly sensitive to the interaction used, so in performing our tests, we just use one interaction for each
type of calculation: PKO1 for RHF-RPA and DD-ME2 for RH-RPA \cite{LGM09}.  The corresponding values for $\delta_C$,
were taken from Table I in Ref.~\cite{LGM09}, and are reproduced here in columns eight and nine of Table~\ref{t:dc}.
Note that the authors of Ref.~\cite{LGM09} only calculated $\delta_C$ values for 8 of the 13 well known superallowed
transitions.

\subsection{Isovector Monopole Resonance (IVMR)}
\label{ss:IVMR}

In 2009, Auerbach \cite{Au09} introduced a model in which he assumed that isospin-symmetry breaking in superallowed
$\beta$ decay is due entirely to mixing with the giant monopole state.

The isovector part of the Coulomb interaction, which appears in Eq.~(\ref{Vc1}), is defined to be the isovector
monopole operator $M_0^{(1)}$; thus
\be
M_0^{(1)} = \sum_i r_i^2 t_z(i),
\label{M0}
\ee
where $M_0^{(1)}$ is a spherical tensor in isospin space of rank 1, with its $z$-component equal to 0.  We write
$|M\rangle$ to be the giant monopole state, which is created by the application of operator $M_0^{(1)}$ to the ground
state.  If the ground state has $N$=$Z$ with isospin quantum numbers $T$=\,0 and $T_z$=\,0, then the giant monopole
state is a unique state with quantum numbers $T$=\,1 and $T_z$=\,0.  But if the ground state has a neutron excess, with
$T$=\,$T_z$=\,$\sfrac{1}{2}(N-Z)$, then the monopole state is split into two components, one with isospin $T$ and the
other with isospin $T$+1.  In this case the ground-state wave function is designated by $|T,T\rangle$ and the two
components of the monopole state by $|M_{T,T}\rangle$ and $|M_{T+1,T}\rangle$.  Furthermore, the isobaric analogue of
this ground state, $|T,T-$1$\rangle$, has its giant monopole state split into three isospin components, $|M_{T-1,T-1}\rangle$,
$|M_{T,T-1}\rangle$ and $|M_{T+1,T-1}\rangle$.

By assuming that the giant monopole state is the sole source of isospin-symmetry breaking in superallowed decays,
Auerbach \cite{Au09} could write the wave functions for the two states involved in the $\beta$ decay as
\bea
|\Psi_1\rangle & = & \Biglb ( |T,T\rangle + \epsilon_0 |M_{T,T} \rangle +
\epsilon_1 |M_{T+1,T}\rangle \Bigrb ) N_1^{-1}
\nonumber \\[2mm] 
|\Psi_2\rangle & = & \Biglb ( |T,T-1\rangle + \eta_{-1} |M_{T-1,T-1} \rangle \Bigrb .
\nonumber \\
& & ~ \Biglb . + \eta_0 |M_{T,T-1} \rangle+ \eta_1 |M_{T+1,T-1}\rangle \Bigrb ) N_2^{-1},
\label{Wavefns}
\eea
where
\bea 
N_1 & = & (1 + \epsilon_0^2 + \epsilon_1^2 )^{1/2}
\nonumber \\
N_2 & = & (1 + \eta_{-1}^2 + \eta_0^2 + \eta_1^2 )^{1/2},
\nonumber \\
\eea
and amplitudes $\epsilon_i$ and $\eta_i$ can be expressed via perturbation theory in terms of Coulomb matrix elements
between the ground state and the respective components of the isovector monopole state.  Based on this result, he
then derived the corresponding isospin-symmetry-breaking correction to superallowed $\beta$ decay, which he wrote
to order $\O (\epsilon^2, \eta^2)$ as
\be
\delta_C = \eta_{-1}^2 + (\epsilon_0 - \eta_0)^2 + \epsilon_1^2 + \eta_1^2
- 2 \epsilon_1 \eta_1 \left ( \frac{2T+1}{T} \right )^{1/2}.
\label{IVMdc}
\ee

Auerbach next argued that the Coulomb matrix elements of differing isospins are all related to each other
via isospin Clebsch-Gordan coefficients.  He thus found that the coefficients $\epsilon_0$, $\eta_{-1}$,
$\eta_0$ and $\eta_1$ could all be expressed in terms of the one isospin-mixing amplitude $\epsilon_1$.  
In this way, the expression for $\delta_C$, Eq.~(\ref{IVMdc}), reduces to
\be
\delta_C = 8(T+1) \frac{V_1}{\xi \hbar \omega A} \epsilon_1^2,
\label{IVMdc1}
\ee
where $\xi$ is related to the particle-hole interaction energy required to place the centroid of the giant
monopole resonance at the appropriate energy; and $V_1$ is related to the strength of the symmetry potential
that sets the energy splitting between the components of the monopole state.  Auerbach chose $V_1 = 100$
MeV, $\xi = 3$, and $\hbar \omega = 41 A^{-1/3}$ MeV and estimated $\epsilon_1^2$ by appealing to a number of
``gross" models discussed in Ref.~\cite{Au83}: a hydrodynamical model, models based on non-energy-weighted
and energy-weighted sum rules, and a microscopic model.  Each enabled him to obtain a simple expression for
$\delta_C$ as a function of the mass number $A$.  As an example, his expression in the microscopic model was
\be
\delta_C = 18.0 \times 10^{-7} A^{5/3}.
\label{IVMdc2}
\ee

We calculated the values of $\delta_C$ from this equation for the thirteen best known superallowed
transitions.  The results, which are listed in column ten of Table~\ref{t:dc}, were used as part of our
comparative tests of all models.

\section{Test results}
\label{s:results}

We have now set the stage for applying the test described in Sect.~\ref{s:test}.  As explained there, our
procedure for each of the six models is to compare that model's set of calculated $\delta_C$ values
(listed in Table~\ref{t:dc}) with the set of values obtained from Eq.~(\ref{finaltest}) and, using the method of least
squares with $\overline{\F t}$ as the adjustable parameter, to optimize the agreement between them.  In effect, the
$\delta_C$ values from Eq.~(\ref{finaltest}) can be thought of as the ``experimental" values: they incorporate the
experimental $ft$ values from Ref.~\cite{HT09}, as well as the small calculated correction terms, $\delta^{\prime}_R$
and $\delta_{NS}$ from Ref.~\cite{TH08} (also listed in Table~\ref{t:dc}).  The parameter, $\overline{\F t}$, is a
normalizer that allows each model to be tested for its success in obtaining a constant $\F t$ value ({\it i.e.} in
agreement with CVC), without regard for whether or not that $\F t$ value ultimately satisfies CKM unitarity.  The
normalized $\chi^2$ for each least-squares fit -- expressed as $\chi^2/n_d$, where $n_d$ is the number of degrees
of freedom -- thus yields a figure of merit for the model used, with smaller $\chi^2/n_d$ values indicating better
agreement.

Although we take the measured $ft$ values from Ref.~\cite{HT09}, strictly speaking the $ft$-value uncertainties
quoted in that reference do not correspond to normal distributions.  Each $ft$ value has three experimental inputs
-- energy, half-life and branching ratio -- and each of these inputs typically includes a number of measurements of
that quantity.  The survey authors adopted the procedures used by the Particle Data Group \cite{PDG08} and, for any
cases in which the measurements when averaged yielded a normalized $\chi^2$ greater than one, they increased the
uncertainty on the average by a scale factor equal to the square root of the normalized $\chi^2$.  This conservative
approach leads to uncertainties on the $ft$ and $\F t$ values that are larger than would be the case for purely
statistical results.  For our present purposes we have set all the scale factors in Ref.~\cite{HT09} equal to 1 and
obtained new uncertainties on the $ft$ values, which are normally distributed (at least to the extent that the
uncertainties assigned by the authors of the original measurements were predominantly statistical).  It is these
re-determined uncertainties that appear in the second column of Table~\ref{t:dc}.

Obviously the uncertainties assigned to the theoretical radiative corrections $\delta^{\prime}_R$, $\delta_{NS}$ and
$\delta_C$ (if any) are not normally distributed statistical quantities.  Therefore, in our first least-squares
test, we used only the re-determined uncertainties for the $ft$ values and no uncertainties at all for any of the
theoretical terms.  The results for $\chi^2/n_d$ appear in the first row below the main body of Table~\ref{t:dc},
labeled ``statistical experimental uncertainties only".  Since this analysis uses only normally distributed uncertainties,
we can proceed to evaluate a confidence level for each model.

We follow the Particle Data Group \cite{PDG08} in defining the confidence level (or ``p-value")
as being
\be
CL = \int_{\chi^2_0}^{\infty}P_{n_d}(\chi^2)d\chi^2,
\label{CL}
\ee
where $P_{n_d}(\chi^2)$ is the $\chi^2$ probability distribution function for $n_d$ degrees of freedom, and $\chi^2_0$
is the value of $\chi^2$ obtained for a particular hypothesis -- in our case, for a particular isospin-symmetry-breaking
model.  With this definition, the confidence level represents the probability that the $\chi^2$ for a valid hypothesis
could exceed the value, $\chi^2_0$, actually obtained for the specific hypothesis being tested.  More loosely, in our
application the confidence level quoted for a particular model can be interpreted as the probability of that model being
a valid one; {\it i.e.} of it being consistent with CVC.   We express each $CL$ as a percent on the next line in the table.

\begin{figure*}[t]
\epsfig{file=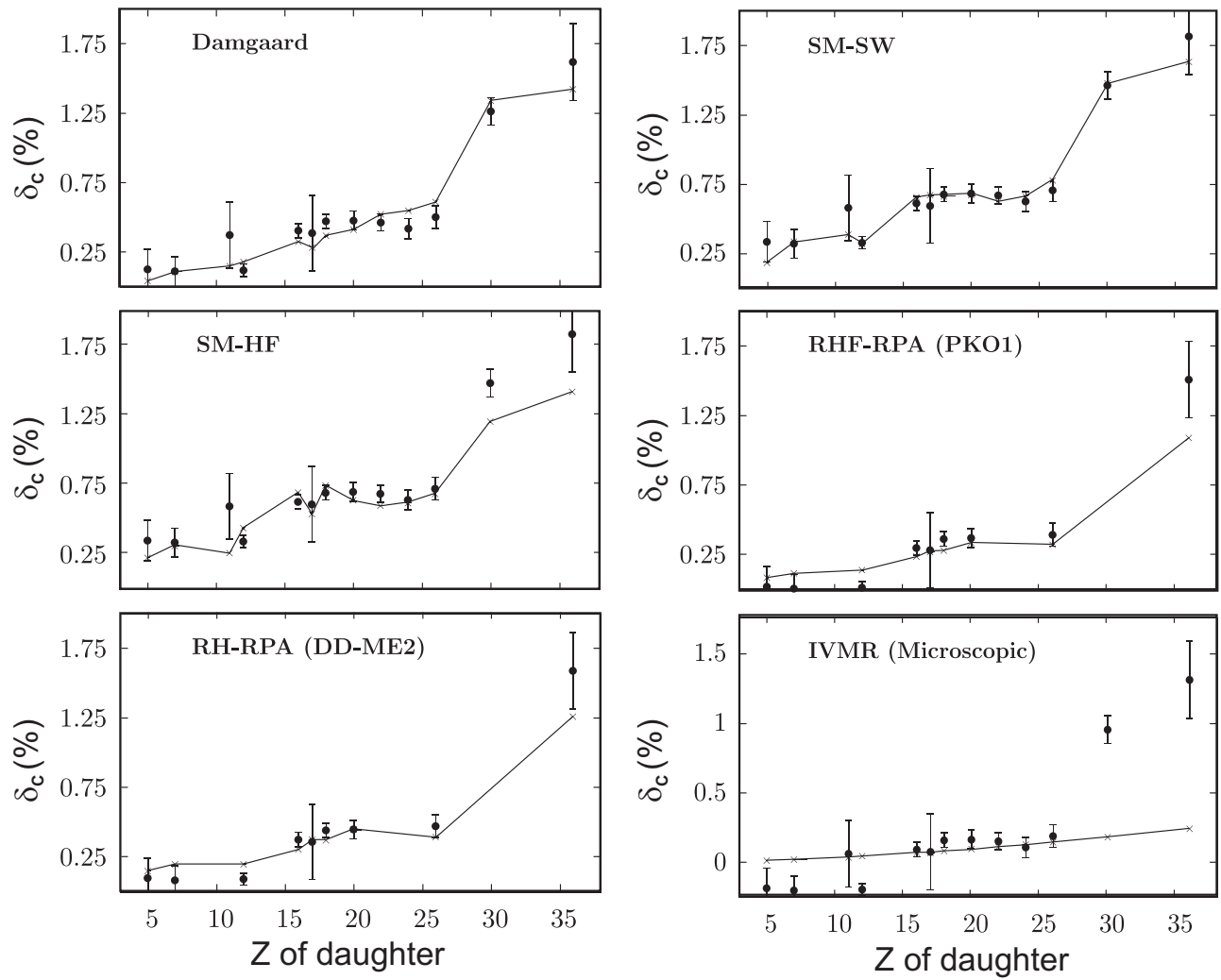,width=17cm}
\caption{Isospin-symmetry-breaking correction, $\delta_C$, in percent units plotted as a function of atomic
number, $Z$, of the daughter nucleus.  The solid circular points with error bars are the values of $\delta_C$
obtained from Eq.~(\ref{finaltest}), with the experimental $ft$ values and the values of $\delta^{\prime}_R$
and $\delta_{NS}$ (and their uncertainties) all taken from Table~\ref{t:dc}.  In effect, we treat these as
the ``experimental" $\delta_C$ values.  The X's joined by lines represent the $\delta_C$ values calculated
by the various models described in the text and identified in the upper left of each graph.  The value of
$\overline{\F t}$ in Eq.~(\ref{finaltest}) has been adjusted in each case by least-squares fitting to optimize
the agreement between the ``experimental" $\delta_C$ values and the calculated ones.  The corresponding values
of $\chi^2/n_d$ are listed in the second-to-last row of Table~\ref{t:dc}.}
\label{fig1}
\end{figure*}

We then present the results of a second least-squares analysis, in which we included uncertainties on the theoretical
radiative corrections, $\delta^{\prime}_R$ and $\delta_{NS}$; of course we retained the re-determined uncertainties already
incorporated for the $ft$ values.  The resulting $\chi^2/n_d$ values appear in the second-to-last line in the table.  These
results are also illustrated in Fig.~\ref{fig1}.

Finally, we list the results from a third least-squares analysis.  Although two of the models, SM-SW and SM-HF, include
theoretical uncertainties on $\delta_C$ values in their original publications, the other four do not; so, to test them
all on an equal footing, we have not used uncertainties on any of the model calculations in our first two analyses.  We
consider this to be the fairest approach.  However, we have also examined what happens to our intercomparison if all
the calculated $\delta_C$ values are assigned the same uncertainties as those originally quoted for the SM-SW
calculations \cite{TH08}.  The values for $\chi^2/n_d$ resulting from this third analysis appear in the bottom line
of Table~\ref{t:dc}.

The most obvious outcome of these analyses is that only one model, SM-SW, produces satisfactory agreement with CVC, 
having $\chi^2/n_d$ = 1.2 and $CL$ = 26\% in the properly statistical analysis.  All of the other 5 models have
confidence levels well below 0.5\%.  Because the two other analyses included non-statistical uncertainties on the
theoretical correction terms in addition to the statistical experimental ones, their values of $\chi^2/n_d$ are
substantially lower, but the relative ranking of the six models is approximately preserved: in all cases the SM-SW
model is by far the best.  It is remarkable that the model which becomes second best when the theoretical
uncertainties are included is the earliest and arguably the most primitive one.  Its success evidently stems from
its treatment of the radial mismatch between the parent and daughter states, which accounts rather well for the
sharp increase in $\delta_C$ between $Z$ = 12 and $Z$ = 16, and between $Z$ = 26 and $Z$ = 30.  It is perhaps
equally striking that the most recent IVMR model fails to reproduce the trend of the data or any of its
characteristic features.

\section{Conclusions}
\label{s:Conc}

Evidently the shell-model with Saxon-Woods radial wave functions, SM-SW, is the only model tested that yields
isospin-symmetry-breaking corrections which, when combined with the experimental $ft$ values, produce $\F t$
values that agree with the CVC hypothesis over the full range of $Z$ values.  This, of course, does not prove
that the SM-SW model is correct in every way; however, it does demonstrate that the other models in their
present form cannot be used to extract a number for $V_{ud}$ and to test CKM unitarity.  As we note in
Sect.~\ref{s:sbd}, if the $\F t$ values are not consistent with one another, then their average has no defined
significance since either the symmetry-breaking model is wrong or CVC itself has failed.

There is a second model, SM-HF, which has many promising features.  As can be appreciated from an examination
of Fig.~\ref{fig1}, its relatively large $\chi^2$ is due to its failure to match the experimental $\delta_C$
values for the cases with $Z \ge 30$.  If we were to restrict ourselves only to the lighter cases, then the model
would agree well with CVC.  This difference at the highest $Z$ values between the SM-SW and SM-HF model calculations
has been known for 15 years, having first been pointed out by Ormand and Brown \cite{OB95} even before the decays of
the highest-$Z$ emitters, $^{62}$Ga and $^{74}$Rb, had yet been precisely measured.  Prompted by the results reported
here, we are currently examining whether this feature of the SM-HF model (as described in Sec.~\ref{ss:SMHF}) is
sensitive to the particular Skyrme interaction used \cite{TH11}.  We have by now sampled 12 different interactions
and have also added a pairing term to the interaction, turning the calculation into a Hartree-Fock-Bogolyubov one.  
However, under no circumstances have we been able to produce agreement with experiment over the full range of $Z$
values.  It is important to realize that both the SM-SW and SM-HF models use identical spectroscopic input, so
it would appear that the high-$Z$ discrepancy is inherent to the SM-HF model itself.  However, it must be admitted
that too little spectroscopic information is known in this region to fully characterize the required model space.
Calculations with larger model spaces and improved Hamiltonians are certainly to be encouraged.       

Fortunately, it is the successful SM-SW model that has principally been employed to calculate the $\delta_C$ values used
in the most recent data survey \cite{HT09}.  As was argued in that survey, the consistency of the $\F t$ values was a
powerful validation of those calculated correction terms and justified the subsequent derivation of $V_{ud}$.  However,
in actually deriving $V_{ud}$ and its uncertainty we incorporated the SM-HF calculations as well, even though we knew
that model had a much poorer $\chi^2$.  Our rationale was one of conservatism.  We enlarged the uncertainty assigned
to the average $\overline{{\cal F}t}$ value to cover both sets of $\delta_C$ calculations in order to be safe by including
some provision for systematic theoretical uncertainties. Whether we continue this practice in future is not yet decided.

For now, though, we know that there are, as yet, no comparably successful competitive models.  More important, we also
have a protocol for testing future models, which is evidently very sensitive to the validity of the model.  Furthermore, even
though it is only the {\em relative} $Z$-dependent variations in $\delta_C$ that are being tested, it would
surely require a pathological fault indeed in the theory to allow the observed nucleus-to-nucleus variations
in $\delta_C$ to be reproduced in such detail while failing to obtain the {\em absolute} values to comparable
precision.

With this perspective, it is now informative to consider the points raised recently by Miller and Schwenk
\cite{MS09}, who claim that the SM-SW model is based on a formally incorrect interpretation of the isospin
ladder operator.  They claim that this ``incorrect" usage must have led to incomplete results for $\delta_C$,
but they do not produce any ``exact" calculations with which to compare.  Instead, they identify a term
involving radial excitations, which they consider to be missing from the SM-SW model, and proceed to evaluate
this term under simplifying assumptions.  They assume that the radial excitations are dominated by mixing with
states having one more radial node ({\it cf.} the Damgaard model) and, further, that the relevant excitations
are dominated by the isovector monopole resonance ({\it cf.} the IVMR model).  Under these conditions they
find that this ``missing" term almost completely cancels the SM-SW-model result and, although they produce no
numbers, they state that this would result in $\delta_C$ values comparable in magnitude to the IVMR-model
results, or even smaller.  Clearly such a result would disagree at least as
strongly with CVC as does the IVMR model.  Therefore if any term is really missing from the SM-SW-model
calculations, the test results presented here show that it must either be independent of $Z$ or else very
small; otherwise the data would become inconsistent with the CVC hypothesis.  Considering that the Coulomb
force is the principal source of isospin symmetry breaking, it is highly unlikely that any large component
of $\delta_C$ could be $Z$ independent.

From an experimental point of view, the results in Fig.~\ref{fig1} clearly demonstrate the importance of
precisely measured $ft$-values.  For example, the very precise values for $^{26}$Al$^m$ (plotted at $Z$ =
12, the atomic number of its daughter) and $^{34}$Cl (see $Z$ = 16) contribute very significantly to
the overall $\chi^2$ for each model fit.  Equally important though are the $ft$ values for transitions that
exhibit large values for $\delta_C$.  The most obvious examples are the decays of $^{62}$Ga (see $Z$ = 30)
and $^{74}$Zr (see $Z$ = 36): their $\delta_C$ values differ enormously from those for the transitions with
$Z$$\leq$54 and this difference plays an important role in differentiating one symmetry-breaking model from
another.  More measurements of both types would be much welcomed in this context.

\acknowledgments

The work of JCH was supported by the U. S. Dept. of Energy under Grant
DE-FG03-93ER40773 and by the Robert A. Welch Foundation under Grant A-1397.
IST would like to thank the Cyclotron Institute of Texas A \& M University
for its hospitality during annual two-month summer visits.

\end{document}